\documentclass[aps,prl,twocolumn,superscriptaddress,showpacs]{revtex4}
\usepackage{graphicx}
\usepackage{amsmath}

\begin{document}


\title{Multi-parameter generalization of nonextensive statistical 
mechanics}

\author{Fabio Sattin}
\email{sattin@igi.pd.cnr.it}

\affiliation{Consorzio RFX, Associazione Euratom-ENEA, Corso Stati Uniti 4, 
35127 Padova, Italy}

\author{Luca Salasnich}
\email{salasnich@mi.infm.it}

\affiliation{Istituto Nazionale per la Fisica della Materia, Unit\`a di 
Milano Universit\`a,  Dipartimento di Fisica, Universit\`a di Milano \\
Via Celoria 16, 20133 Milano, Italy}

\date{\today}

\begin{abstract}
We show that the stochastic interpretation of Tsallis' 
thermostatistics given recently by Beck [Phys. Rev. Lett {\bf 87}, 
180601 (2001)] leads naturally to 
a multi-parameter generalization. The resulting class of distributions 
is able to fit experimental results which cannot be reproduced within 
the Boltzmann's or Tsallis' formalism.
\end{abstract}


\maketitle

Nonextensive statistical mechanics (NESM)  
introduced by Tsallis \cite{tsallis} has gained a considerable interest 
in several fields of physics because of its capability to describe a wealth 
of disparate phenomena (from anomalous diffusion, to turbulent systems, to 
astrophysical systems, etc ...) within a single formalism, generalization 
of the standard statistical--mechanical one with the addition of the single 
free parameter (entropic index) $q$.
Recently it has been shown how to relate $q$  with 
the internal microscopic properties of the system under consideration. This 
has been done by Wilk and Wlodarczyk \cite{wilk}: they have shown that, when $q \geq 1$, 
the NESM canonical distribution $\rho_{q}(H,\beta_{0})$ for the system with 
Hamiltonian $H$ can be written as an average of the 
usual Boltzmann-Gibbs factor over the inverse 
temperature $\beta$:
\begin{equation}
 \label{eq:rhof}
\rho_{q}(H,\beta_{0}) = \int_{0}^{\infty} d\beta \exp(- \beta H) 
f_{q}(\beta,\beta_{0}) \quad ,
\end{equation}
where $f_{q}(\beta,\beta_{0})$ is a weight function whose meaning is that of a  
probability distribution function for $\beta$ which is, therefore, no 
longer a fixed parameter; instead, the macroscopically visible value 
is just its average value $\beta_{0}$.
Fluctuations in $\beta$ are related to coherent fluctuations existing 
in small parts of the system in respect to the whole system, due to the 
existence of long range correlations. \\
Recently, Beck \cite{beck1} has been able to give an 
interpretation of the fluctuating $\beta$ as a function of 
stochastically varying microscopic variables. In order to recover 
Tsallis' results, Beck was forced to impose some constraints over 
$\beta$ or, equivalently, the microscopic dynamics of the system. In 
this paper we show that, following Beck's approach but relaxing 
these constraints, we are able to derive an entire new class of 
distributions,
which reduce to Tsallis' distribution under suitable limits. We will show 
that some members of this class are able to reproduce experimental 
 results which would be outside the reach of Tsallis' 
formalism. 

\begin{figure}
\includegraphics[scale=0.4]{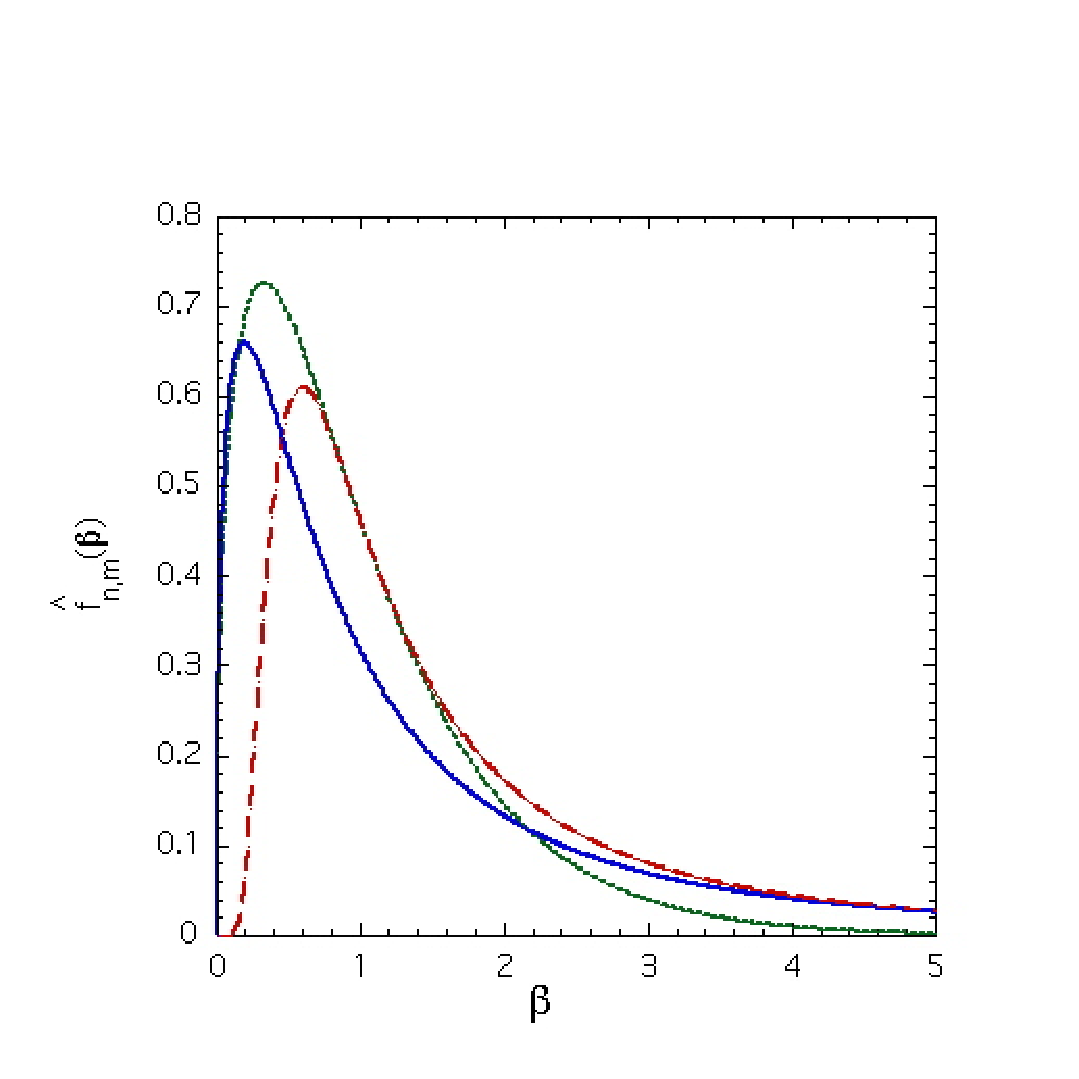}
\caption{\label{fig:zero} 
Probaility distribution $\hat{f}_{n,m}(\beta)$ from Eq. (\ref{eq: PDF}), with $\beta_{0} = 1$.
Solid line, $\hat{f}_{3,3}$; dotted line, $\hat{f}_{3,\infty}$; dashed 
line,  $\hat{f}_{\infty,3}$. }
\end{figure}

\begin{figure}
\includegraphics[scale = 0.4]{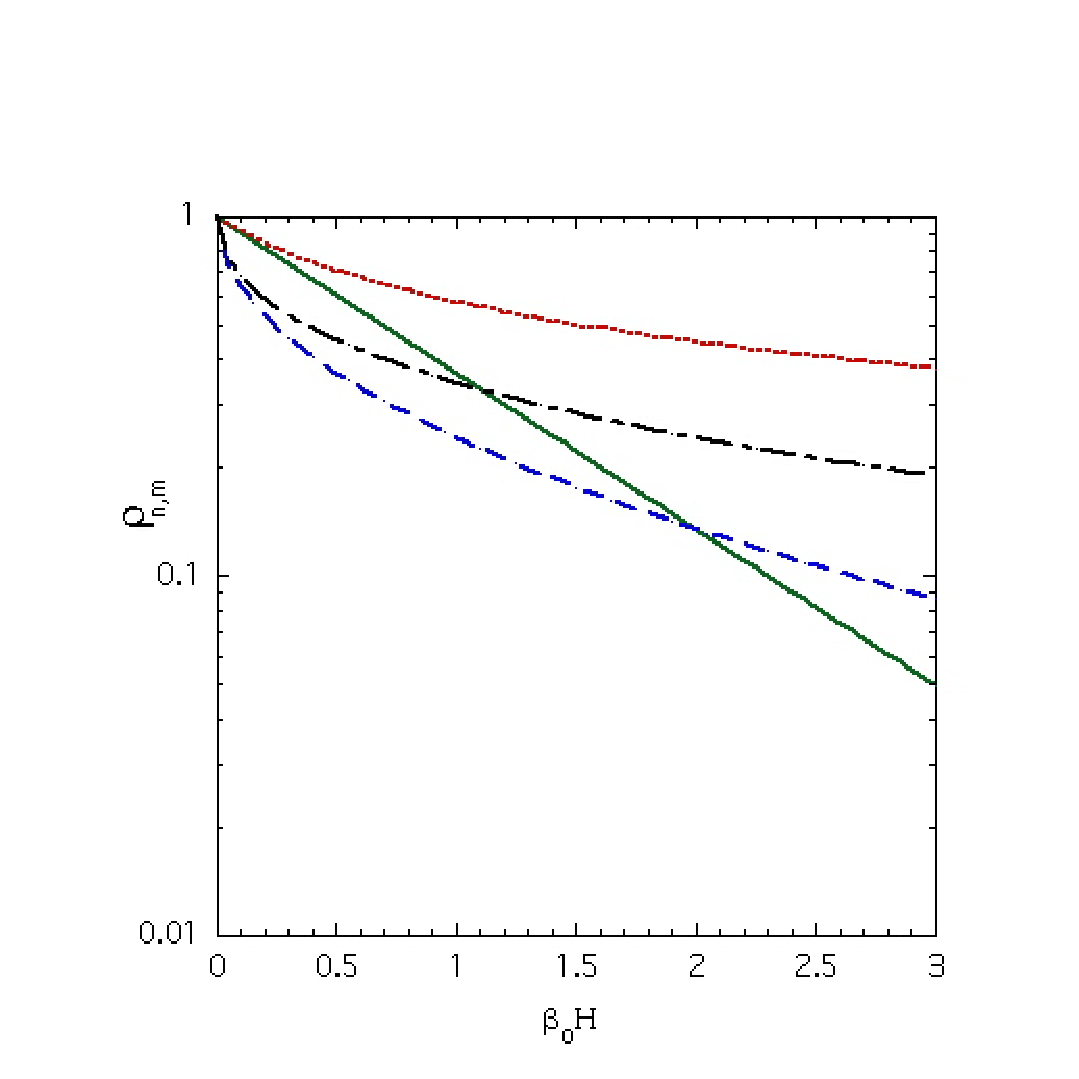}
\caption{\label{fig:uno} Generalized canonical distributions 
$\rho_{n,m}$ as a function of the scaled energy $\beta_{0} H$.   
Solid line, $ n \to \infty , m \to \infty$ [this yields the usual 
Boltzmann-Gibbs (BG) case $\exp(-\beta_{0} H)$]; 
dotted line, Eq. (\ref{eq:rhon}) with $ n = 1$ , corresponding to the Tsallis distribution with $ 
q = 2$); dashed line, Eq. (\ref{eq:rhom}) with $ m = 1$ ; 
dotted-dashed line, $ n = 2, m = 1$. }
\end{figure}

\begin{figure}
\includegraphics[scale = 0.4]{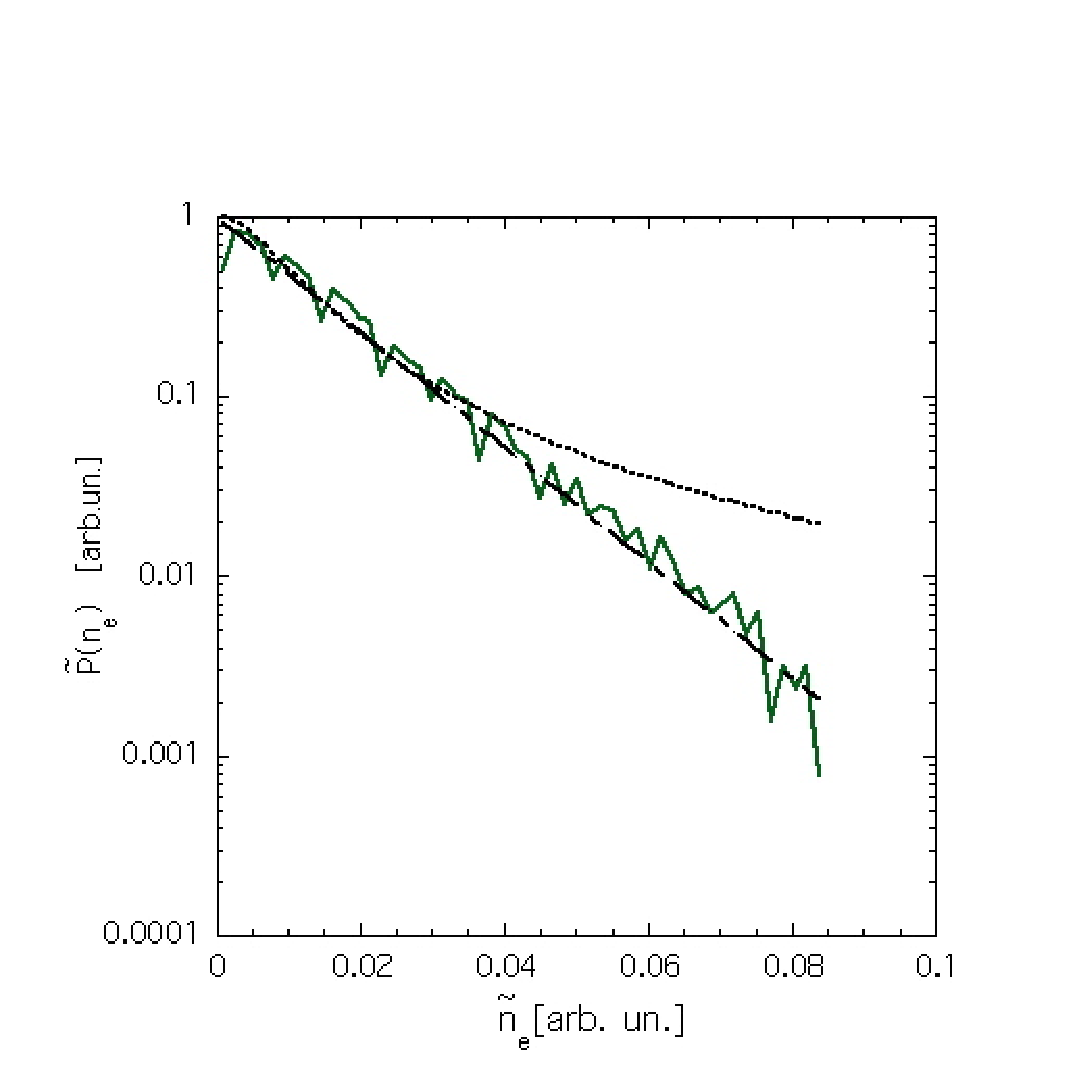}
\caption{\label{fig:due} Probability distribution $P(\tilde{n}_{e})$ 
of the electronic density fluctuations $\tilde{n}_{e}$.
Broken line, experimental data from ref. \cite{antar} (only the side of
positive fluctuations is shown); dotted line, best fit 
using Tsallis' distribution (\ref{eq:pn}); dashed line, best fit with 
curve (\ref{eq:pm}) and $m = 1$. }
\end{figure}

To start with, we quote the same example used in Beck's paper: 
let us set $ H = u^2/2$ and suppose that 
the generalized velocity $u$ satisfies the Langevin equation
\begin{equation}
\dot{u} = - \gamma u + \sigma L(t) 
\end{equation}
with $ L(t) $ Gaussian white noise of unit amplitude, $\sigma$ strength of 
the noise, $\gamma$ friction coefficient. This is the Brownian particle problem \cite{kampen}.
For this case, it can be shown that the temperature 
$1/\beta$ is related to the microscopic parameters $\gamma, \sigma$ by
\begin{equation}
\label{eq:caso1}
\beta = \gamma/\sigma^2 \quad .
\end{equation}
Beck shows that Tsallis's distribution can be recovered if $\beta$ is 
characterized by a $\chi^2$ distribution with $n$ degrees of freedom 
\cite{statistica}
\begin{equation}
\label{eq:chi2}    
\hat{f}_{n}(\beta,\beta_{0}) = {\left( {n \over 2}\right)^{n \over 2} \over 
\beta_0 \Gamma\left({n \over 2}\right) } \left({\beta \over \beta_0} 
\right)^{{n \over 2} -1} \exp\left(- { n 
\beta \over 2 \beta_0} \right)  \quad ,
\end{equation}
where $f_{q}(\beta, \beta_{0}) = \hat{f}_{n}(\beta, \beta_{0})$ 
provided that $ q = 1 + 2/(n + 1) $.
Such a distribution arises if $\beta$ can be written as 
a sum of stochastic variables: 
\begin{equation}
\label{eq:somma}
\beta = \sum_{i=1}^{n} X_i^2 \quad , 
\end{equation}
with $ <X_{i}> = 0 $ and $<X_i^2> = \beta_0 /n$, so that $<\beta> = 
\beta_{0}$ and $<\beta^{2}> - <\beta^{2}> =  \beta_{0}^{2}\, (2/n)$. The $\chi^{2}$ 
distribution is a common distribution, occurring in many physical 
problems, and is central in the problem of estimating parameters from 
data \cite{roe}. \\
Some points are worth stressing at this stage: \\ 
(i) The macroscopic parameter $\beta$ is written in terms of other 
parameters more directly related to the microscopical dynamics of the 
system at hand, just as in Eq. (\ref{eq:caso1}). We just mention another example:
in the study of fully developed turbulence,  where $u$ is a local 
velocity difference, $\beta = (\varepsilon \tau)^{-1} $,
with $\varepsilon$ spatially averaged energy dissipation rate and $\tau$ 
typical time for the energy transfer. \\
(ii) It is obvious that, if $\beta$ is a stochastic variable, {\it a 
fortiori} the microscopic quantities $\gamma, \sigma, \ldots$ must 
also be  stochastic variables, therefore characterized by their own 
probability distribution functions (PDFs).
(iii) Relations of the kind (\ref{eq:somma}) impose severe 
constraints upon the PDFs of the microscopic variables. 
For example, to recover Eq. (\ref{eq:chi2}) starting from Eq. 
(\ref{eq:caso1}) there is the trivial choice: $\gamma$  
$\chi^2$-distributed and $\sigma^2$ a constant; 
it is difficult (and perhaps impossible) to devise other distributions which 
lead to Eq. (\ref{eq:chi2}). \\
The main idea of this paper is that if $\beta$ is a function  
of some more fundamental  
stochastic control variables, then the 
by far more logical path is the following: to guess statistical 
distributions for the microscopic quantities and, from them, to work out the 
corresponding distribution for $\beta$. Since $\beta$ may have 
infinite functional dependences from microscopic variables, we can 
expect the PDF of $\beta$  to have a large range of analytical forms,  
depending on a large number of parameters [we expect as many of them as the number 
of microscopic variables that control $\beta = \beta(Y,Z, \ldots) $].  \\
Some simple rules, however, still allow to drastically reduce the class of 
likely distributions. First, although the  PDF for each of the variables $Y, Z, 
\ldots$ , may be arbitrary, the same reasoning of Eqns 
(\ref{eq:chi2},\ref{eq:somma}) still holds: that is, the $\chi^{2}$ 
distribution for each variable is a very convenient choice. For 
example, the $\chi^{2}$ distribution can tranform into a delta 
distribution, thus allowing for 
well deterministic, non--stochastic quantities in the limit $n \to 
\infty$.  Hence, we will suppose all the stochastic variables to be 
$\chi^{2}$-distributed, possibly with different degrees of freedom.
In second place, a simplicity principle suggests that the most frequently occurring cases  
should be those where $\beta$ is some simple combination of a small 
number of variables. Some examples are given in the above 
expressions (e.g., Eq. \ref{eq:caso1}). 
The simplest function of all is the sum of stochastic variables: 
$\beta = Y + Z + ... $ . However, with the previous choice for the 
  PDFs of $Y, Z, \ldots$, it is possible to show that it is a trivial 
case, since it reduces to a $\chi^{2}$ distribution \cite{statistica}.
The next nontrivial cases, thus, are those involving products and 
ratios of one or two control variables: $ Y \times Z $, $ Y/Z$ , $1/(Y 
Z), \ldots $.\\
Our aim now is to compute a few examples of PDFs of $\beta$ 
and to compare the results with the Tsallis' formalism. We 
will do the computation for the case of $\beta$ ratio of two stochastic 
variables: $\beta = Y/Z$. This is particularly convenient since: (i) it 
generalizes the example given by Beck (Eq. \ref{eq:caso1}) ; (ii) it 
is a particular case of $\beta = 1/(\varepsilon \tau)$, when $Y$ and either 
$\varepsilon$ or $\tau$ are constants. \\
The probability distribution function for the two $\chi^{2}$ independent variables 
$Y$, $ Z$ of degree $n$, $m$ respectively, is given by
\begin{eqnarray}
\label{eq:prodotto}
\nonumber
&\hat{f}_{n}&(Y,Y_{0}) \hat{f}_{m}(Z, Z_{0})  = { \left({ n \over 2 Y_0}\right)^{n \over 2} 
\left( {m \over 2 Z_0} \right)^{m \over 2} 
\over \Gamma\left({n \over 2}\right) \Gamma\left({m \over 
2}\right) } \\
&\times& Y^{{n \over 2} -1} Z^{{m \over 2} -1}   
 \exp\left( - { n Y \over 2 Y_0} \right)  \exp\left( - { m Z \over 2 Z_0} \right)  \quad .
\end{eqnarray}
($\Gamma$ is the factorial function $\Gamma(z) = \int_{0}^{\infty} 
t^{z-1} e^{-t} d\,t$).
We set $ \beta = Y/Z$, $\beta_0 = Y_0/Z_0$ and regard $\beta$ 
and $Z$ as independent variables; after integration over $Z$, we get
\begin{equation}
\label{eq: PDF}	
\hat{f}_{n,m}(\beta, \beta_{0})  = {\Gamma\left({n + m \over 2}\right) \over \Gamma\left({n 
\over 2}\right) \Gamma\left({ m \over 2}\right)} \left({ n \over 
m}\right)^{n \over 2} { \left(\beta/\beta_0\right)^{{ n \over 2} -1} \over \left[ 1 + {n 
\over m} {\beta \over \beta_0} \right]^{n + m \over 2} } {1 
\over \beta_0} \quad ,
\end{equation}
which is known as F distribution in statistics. This is the 
main result of the work, since the statistical properties of the 
system are determined through the two-parameters canonical distribution, 
generalization of Eq. (\ref{eq:rhof}):
\begin{equation}
\label{eq:rhonostra}    
\rho_{n,m}(H,\beta_{0}) = \int_{0}^{\infty} d\beta \exp(-\beta H) 
f_{n,m}(\beta,\beta_{0}) \quad .    
\end{equation}    
The main feature of Eq. (\ref{eq: PDF}) is that the exponential term of 
Eq. (\ref{eq:chi2}) has disappeared, replaced by a power-law term. 
One should expect this term to depress high-energy tails 
in Eq. (\ref{eq:rhonostra}). In order to have an insight on the 
trends of Eq. (\ref{eq: PDF}), let 
us consider some interesting limits. First of all, we observe that, in 
the limit $ m \to \infty$,
\begin{equation}
\label{eq:fn}
\hat{f}_{n,\infty}(\beta, \beta_{0}) = {\left( {n \over 2}\right)^{n \over 2} \over 
\beta_0 \Gamma\left({n \over 2}\right) } \left({\beta \over \beta_0} 
\right)^{{n \over 2} -1} \exp\left(- { n 
\beta \over 2 \beta_0} \right)  \; .
\end{equation}
We recover the $\chi^2$ distribution (Eq. \ref{eq:chi2}) since, 
in the limit of infinite degrees of freedom, the distribution for $Z$ shrinks to a delta 
distribution, so we are actually dealing with just one stochastic 
variable, $Y$.
It is completely new  the limit $ n \to \infty$ (that is, we are computing the  PDF of the variable $1/Z$), for which we get:
\begin{equation}
\label{eq:fm}	
\hat{f}_{\infty,m}(\beta, \beta_{0})  = {\left({m \over 2}\right)^{m \over 2} \over \beta_{0} \Gamma\left({m \over 2}\right)} 
\left( { \beta_0 \over \beta}\right)^{{m \over 2} +1}  \exp\left( - { m \beta_0 \over 2 \beta} \right) \; .
\end{equation}  
In order to give a 
visual insight, we plot in Fig. \ref{fig:zero} some examples of these 
distributions. The qualitative shape is rather similar.
The occupations factors are computed through eq. 
(\ref{eq:rhonostra}). We give explicit 
expressions for the cases corresponding to the two limits $n \to 
\infty$, $m \to \infty $:
\begin{subequations}
\label{eq:rhos}    
\begin{eqnarray}
&\rho_{n, \infty}(H,\beta_{0})& = 
{1 \over \left[1 + {2 \over n} \beta_{0} H \right]^{n \over 2} } 
\quad , \label{eq:rhon} \\
&\rho_{\infty,m}(H, \beta_{0})& = { \left( 2 m \beta_{0} H\right)^{m \over 4} 
\over 2^{{m \over 2} -1} \Gamma\left({m \over 2}\right) } K_{m \over 
2}\left[\sqrt{ 2 m \beta_{0} H} \right] \; , 
\label{eq:rhom}
\end{eqnarray}
\end{subequations}
where $K$ is the modified Bessel function of order $m/2$ and erfc is 
the complementary error function: erfc =  1 - erf.
The general case of arbitrary $n , m$ can be explicitly written down, but it is not 
revealing since it involves complex combinations of 
hypergeometric function, difficult to visualize.
We plot in Fig. \ref{fig:uno} the standard Boltzmann--Gibbs 
factor together with the curves (\ref{eq:rhos}). 
In general, the new distributions are characterized by tails intermediate between 
Boltzmann's and Tsallis' statistics.
We can obtain the probability distribution $P_{n,m}(u)$ for the 
generalized velocity $u$ once an explicit form for $H = H(u)$ is given. 
By assuming the usual form $ H = u^2 /2$: 
\begin{eqnarray}
    P_{n,\infty}(u) &=& \sqrt{\beta_{0} \over 2 \pi} {\Gamma\left( {n 
\over 2}\right) \over \Gamma\left({ n - 1 \over 2}\right) } {1 \over 
\left[ 1 + {\beta_{0} \over n} u^{2} \right]^{n\over 2} } \quad , 
\label{eq:pn}\\
P_{\infty,m}(u) &=&  {\beta_{0}^{m + 2 \over 4} |u|^{m\over 2} \over 2^{m -2 \over 
4} \pi^{1\over 2} \Gamma\left({ m + 1 \over 2}\right) } 
K_{m\over 2}\left[\sqrt{m \beta_{0}} |u|\right] \; . \label{eq:pm}
\end{eqnarray}
Notice that the function $K$ yields a typical $ \exp( -c H^{1\over 
2})$ or $\exp(- c' |u|)$ dependence. Such a law 
cannot be recovered within the Tsallis' formalism, which predicts 
power-law dependences. Therefore we take it as a signature of this new 
class of functions. It may be of interest to notice that the dependence 
on $|u|$ comes from the variable at the denominator of $\beta$, while 
the numerator provides a dependence on $u^2$. In the general case, 
both $|u|$ and $u^{2}$ terms do appear. \\
The question arises obviously if such distributions do exist in nature. 
We are interested in fluctuations of some quantity: for 
independent fluctuations, the Central Limit Theorem predicts a Gaussian 
PDF. If departures from Gaussianity are described 
in terms of Tsallis' statistics, only   PDFs with power--law 
asymptotics may be included. On the basis of what told before, we must 
look for PDFs with exponential tails.
Actually, in literature are presented several examples of quantities 
whose   PDFs are (at least on some ranges) exponential. 
We briefly mention the numerical computation of the velocity distribution function 
solution of the Enskog--Boltzmann equation for a granular gas \cite{brey};
other hints come from calculations of the large-scale probability 
density distribution in astrophysics \cite{Gaztanage} and from the 
numerical simulation of stresses in sheared granular materials 
\cite{barden}. A field where several well documented examples can be found is 
the study of turbulence in fluids. We refer in particular to papers 
\cite{castaing,vincent,min,chavanis}). The 
quantity we are interested in here is the  PDF of the velocity difference 
between two spatial points. It is found both experimentally and 
numerically that this quantity shows an exponential tail. In particular, 
in paper \cite{castaing} the departure from a Gaussian form is 
interpreted within a formalism very close to ours, where the average 
(\ref{eq:rhonostra}) is done using  their equivalent of 
$\hat{f}_{n,m}(\beta,\beta_{0})$ given by a log-normal function 
[see their equations (3.1-3.4)]. 
The paper \cite{min}, furthermore, shows that the tails of this  PDF 
can smoothly vary between the Cauchy form (which is a particular kind of Tsallis' 
distribution) to Gaussian form passing through the exponential form, by varying 
a few control parameters. This is strikingly reminiscent of varying 
$n,m$ parameters in our formalism. \\
In more detail, we can quote two experimental studies from fusion plasma physics: 
in the  first paper \cite{antar} it is presented a study of the density fluctuations 
existing in a thermonuclear fusion device. The time behavior of the 
electron density $n_{e}$ close to the boundary of the device was measured 
with high sampling frequency, thus allowing to compute the  PDF of the 
fluctuation $\tilde{n}_{e} = n_{e} - <n_{e}>$. It was found that the curve is highly asymmetrical, with 
the negative wing approximately gaussian, and the positive one nearly 
exponential. In Fig. \ref{fig:due} we fit the experimental data with both Tsallis' and our 
curve, showing that the former curve cannot fit the tail of the 
experimental distribution. 
\begin{figure}
\includegraphics[scale = 0.4]{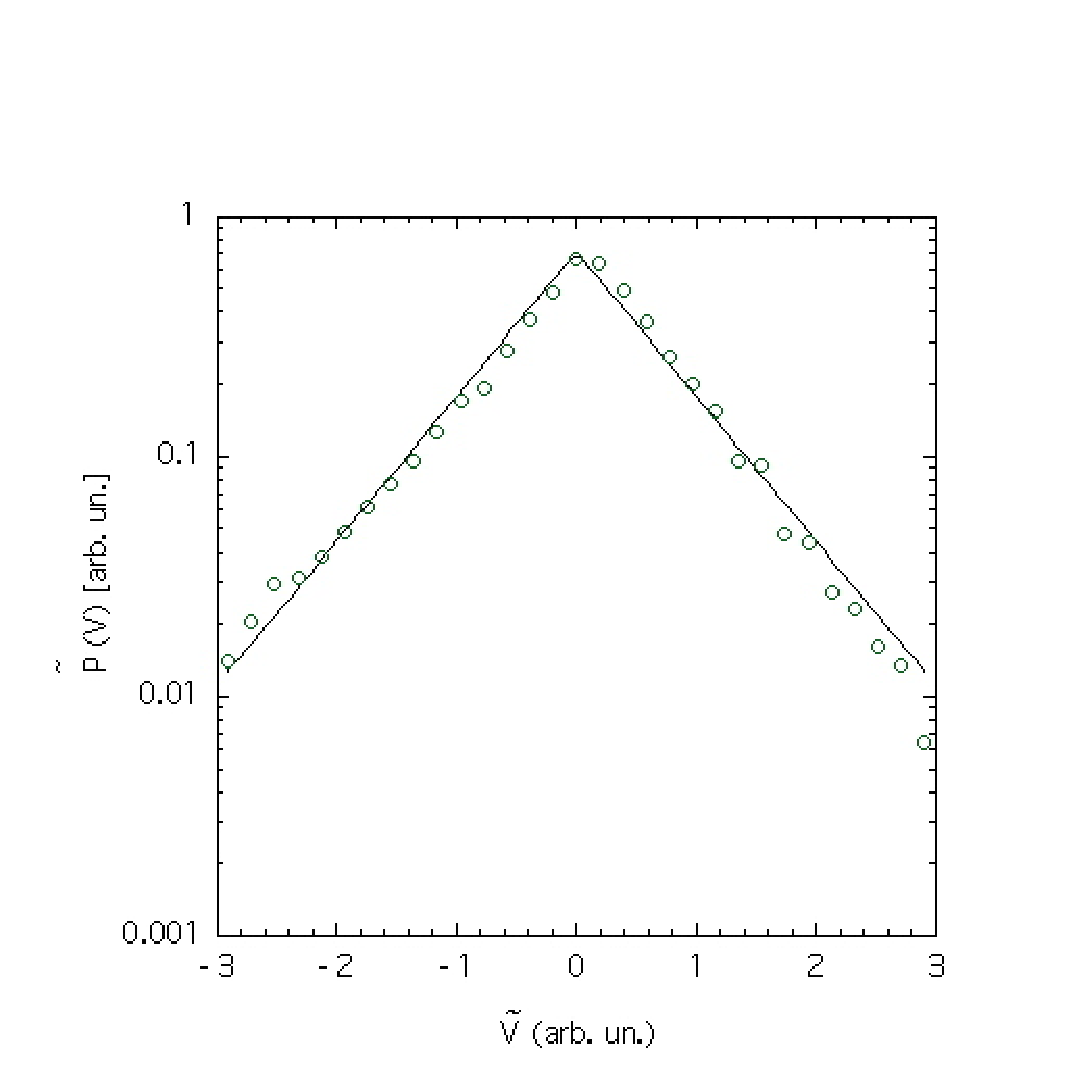}
\caption{\label{fig:tre} Probability distribution $P(\tilde{V})$ 
of the electrostatic potential fluctuations $\tilde{V}$.
Circles, experimental data from ref. \cite{carbone}; solid line, best fit with 
curve (\ref{eq:pm}) and $m = 1$. }
\end{figure}
Rather closely related, we mention a second paper, dealing with a statistical analysis of 
electrostatic potential fluctuations, still in the edge of a plasma 
\cite{carbone}. A wavelet analysis of the data allowed there to compute PDFs as function of 
the time scale of the fluctuations. A scaling law for PDFs was 
recovered by fitting them with stretched exponentials: $P(X) \approx 
\exp \left(-b|X|^{\alpha}\right)$. The parameter $\alpha$ is function 
of the time scale, varying between 1  (exponential distribution) and 2
(Gaussian distribution). In Fig. \ref{fig:tre} the case closest to an 
exponential is shown. 

We think we have given in this work constructive evidence of the existence of 
generalized nonextensive distributions. The very simple   PDFs we 
have computed, seemingly gave us the tools to describe 
complicated phenomena. \\
A crucial point is the choice of 
the microscopic variables, since one could always choose varying 
definitions for them so as to identify several different cases within the
same classes of functions. Therefore, work in this direction should: (i) either show 
that trivial redefinitions of variables are not important for the 
final result, (ii) or find that some sets of variables are preferred 
with respect to all the others.   

\begin{acknowledgments}
We gratefully acknowledge G. Antar for providing us with the 
experimental data of Fig. \ref{fig:due}, E. Martines for the data 
used in Fig. \ref{fig:tre}.
\end{acknowledgments}



\end{document}